\documentclass[reprint,twocolumn, superscriptaddress,
 amsmath,amssymb,
 aps,
 longbibliography,
 footinbib,
]{revtex4-2}

\usepackage{graphicx}
\usepackage{dcolumn}
\usepackage{bm}
\usepackage[dvipsnames]{xcolor}
\usepackage{comment}
\usepackage{fontenc}
\usepackage{float}
\usepackage[colorlinks=true,bookmarks=false,citecolor=blue,urlcolor=blue]{hyperref}

\usepackage[utf8]{inputenc}

\begin{document}
\title{Subwavelength Raman Laser Driven by Quasi Bound State in the Continuum}

\author{Daniil Riabov}
\affiliation{ITMO University, Department of Physics and Engineering, Saint-Petersburg, Russia}

\author{Ruslan Gladkov}
\affiliation{ITMO University, Department of Physics and Engineering, Saint-Petersburg, Russia}

\author{Olesia Pashina}
\affiliation{ITMO University, Department of Physics and Engineering, Saint-Petersburg, Russia}

\author{Andrey Bogdanov}
\affiliation{ITMO University, Department of Physics and Engineering, Saint-Petersburg, Russia}
\affiliation{Qingdao Innovation and Development Center, Harbin Engineering University, Qingdao 266000, Shandong, China}

\author{Sergey Makarov*}
\affiliation{ITMO University, Department of Physics and Engineering, Saint-Petersburg, Russia}
\affiliation{Qingdao Innovation and Development Center, Harbin Engineering University, Qingdao 266000, Shandong, China}

\begin{abstract}
* Corresponding authors: s.makarov@metalab.ifmo.ru; daniil.ryabov@metalab.ifmo.ru

Raman lasers is an actively developing field of nonlinear optics aiming to create efficient frequency converters and various optical sensors. Due to the growing importance of ultracompact chip-scale technologies, there is a constant demand for optical devices miniaturization, however, the development of a nanoscale Raman laser remains a challenging endeavor. In this work, we propose a fully subwavelength Raman laser operating in visible range based on a gallium phosphide nanocylinder resonator supporting a quasi bound state in the continuum (quasi-BIC). We perform precise spectral matching of nanoparticle's high-$Q$ modes with the pump and detuned Raman emission wavelengths. As a result of our simulations, we demonstrate a design of Raman nanolaser, ready for experimental realization, with the lasing threshold expected to be as low as $P_{\mathrm{th}} \approx 21~\mathrm{mW}$. The suggested configuration, to the best of our knowledge, represents the very first prototype of a low-threshold Raman nanolaser with all the dimensions smaller than the operational wavelength.

\textbf{KEYWORDS}: Raman lasing, nonlinear nanophotonics, quasi-BIC, low threshold, nanoparticle.

\end{abstract}
\maketitle

\section{Introduction}

Subwavelength sources of coherent light are usually called \textit{'nanolasers'}~\cite{gu2017semiconductor,ning2019semiconductor}. The small sizes of the nanolasers make them a versatile tool for such applications as advanced telecommunication, sensing and biomedical technologies~\cite{ma2019applications}. Raman nano- and micro-lasers is a separate family of lasers, allowing not only to convert wavelength but also use them for high-sensitivity detection~\cite{nishizawa1980semiconductor,rong2005continuous,rong2005all,latawiec2015chip}. Recently, bound states in the continuum (BIC) concept~\cite{hsu2016bound} offered a novel type of localized high-$Q$ modes in single nanoparticles or subwavelength resonators (quasi-BIC)~\cite{rybin2017high,koshelev2020subwavelength}, which was successfully used for semiconductor luminescent nanolasers~\cite{mylnikov2020lasing}. In general, enhancement of nonlinear effects requires optimization for various parameters of optical cavities to provide not just high quality factors ($Q$), but besides, precise modes spectral matching with both pump and emission wavelengths~\cite{celebrano2015mode, cowan2002mode}. Raman-active materials can contribute to high values of the gain, but spectral proximity of pump and emission bands as well as extremely narrow lines of the gain profile in semiconductors impose additional challenges to optimization of Raman lasers performance~\cite{kippenberg2004ultralow,agarwal2019nanocavity, zograf2020stimulated}.

Raman emission from semiconductors is a process involving the inelastic scattering of photons by crystalline vibrations -- optical phonons. When incident on the Raman active medium a photon could either emit (Stokes component) or absorb (anti-Stokes) a phonon via electron-phonon interaction which results in a spectral shift of the scattered photon~\cite{cardona2005fundamentals}. The value of this shift is defined by the optical phonon energy and is a unique fingerprint of each semiconductor. Here we focus on the Stokes component of Raman emission since this process has a larger probability in thermal equilibrium systems due to Boltzmann statistics and, consequently, results in higher intensity of the output signal as compared to the anti-Stokes peak~\cite{cardona2005fundamentals}. 
\begin{figure}[ht!]
\includegraphics[width=\linewidth]{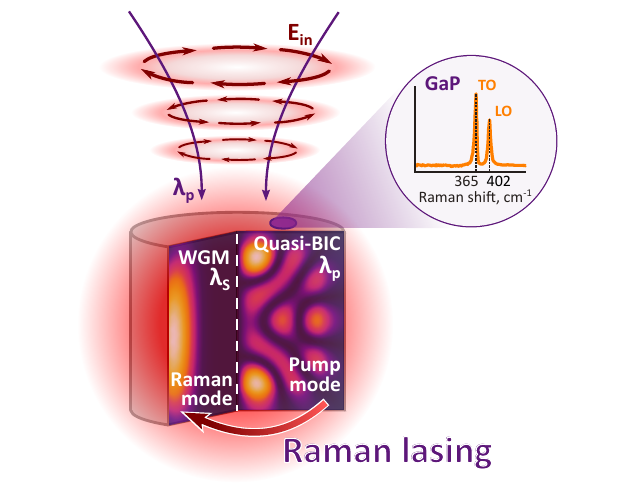}
\caption{\textbf{Conceptual design of Raman laser on quasi-BIC.} GaP nanocylinder is irradiated by a focused azimuthal vector beam exciting quasi-BIC mode at the pump wavelength $\lambda_{\mathrm{p}}$ and another high-Q whispering gallery mode at the Stokes wavelength $\lambda_{\mathrm{S}}$ shifted from the pump due to inelastic scattering on optical phonons as shown in the inset~\cite{Xiong:2003:1533-4880:335}.    \label{fig: concept}}
\end{figure} 

In this work, we theoretically propose a subwavelength Raman laser based on the quasi-BIC in a single cylindrical nanoparticle made of a gallium phosphide (GaP) and located on a sapphire substrate. Since Raman scattering is a nonlinear process involving electromagnetic interaction at two different wavelengths, we start our search of the optimal design from general considerations of the process efficiency. Namely, we finely tune the geometrical parameters of the optical resonator in a way to precisely adjust the position of eigenmodes at the excitation wavelength $\lambda_{\mathrm{p}}$ (pump mode) and at the corresponding narrow line of Stokes Raman emission with a central wavelength $\lambda_{\mathrm{S}}$ (Raman mode)~[Fig.~\ref{fig: concept}]. Moreover, since the enhancement of stimulated Raman emission is defined by the effective light-matter interaction length $L\approx Q\lambda/2\pi n$,  where $n$ is a medium refractive index~\cite{lin1997microcavity}, we can effectively expand active medium by increasing emission mode quality factor, while keeping resonator volume sufficiently small. For this purpose, we utilize high-$Q$ whispering gallery mode (WGM) as the Raman mode at the emission wavelength $\lambda_{\mathrm{S}}$.

We choose lasing threshold as the main parameter for designs efficiency comparison. Therefore, we develop a rigorous model for threshold calculation based on the general wave theory for stimulated Raman scattering. However, given the demanding computational requirements of such a method, we narrow down the search scope of the suitable designs by exploiting a simplified analytical model for Raman lasing. This model gives an estimation for the input power and allows to choose a design providing the lowest lasing threshold.

Finally, we use our model to propose a particular experimental realization of a Raman nanolaser. We choose $\lambda_{\mathrm{p}} = 633$~nm as the pump wavelength that corresponds to HeNe (standard Raman laser). Also, we choose GaP as material for the resonator, since it possesses virtually zero optical losses around $\lambda_{\mathrm{p}} = 633$~nm~\cite{khmelevskaia2021directly} and provides high Raman gain~\cite{wu2009cavity} being comparable to that for bulk silicon~\cite{sirleto2012giant}, which is a common Raman active medium. Besides, Raman gain of GaP profile has two peaks, which provides us with an additional degree of freedom while searching for the resonant modes. Finally, established technologies exist to create high-quality GaP structures by using electron beam lithography~\cite{wilson2020integrated, anthur2020continuous, khmelevskaia2021directly}.

\section{Results}
\subsection{Optical modes optimization for excitation and emission}

The intensity of the Raman signal is proportional to the optical confinement factor of the mode at $\lambda_{\mathrm{p}}$, therefore, one needs to better localize the electromagnetic field at the incident wavelength to decrease Raman lasing threshold. To enhance light trapping efficiency, we apply the concept of quasi-BICs in the single optical resonator~\cite{rybin2017high}. Strong suppression of radiative losses for such states leads to an increase in the $Q$ factor and strong enhancement of the local field. Nevertheless, quasi-BICs are still accessible from the radiation continuum which allows one to excite them significantly easier than other high-order modes with large quality factors~\cite{cai2020overview}. 

\begin{figure*}[ht!]

\includegraphics[width=0.99\linewidth]{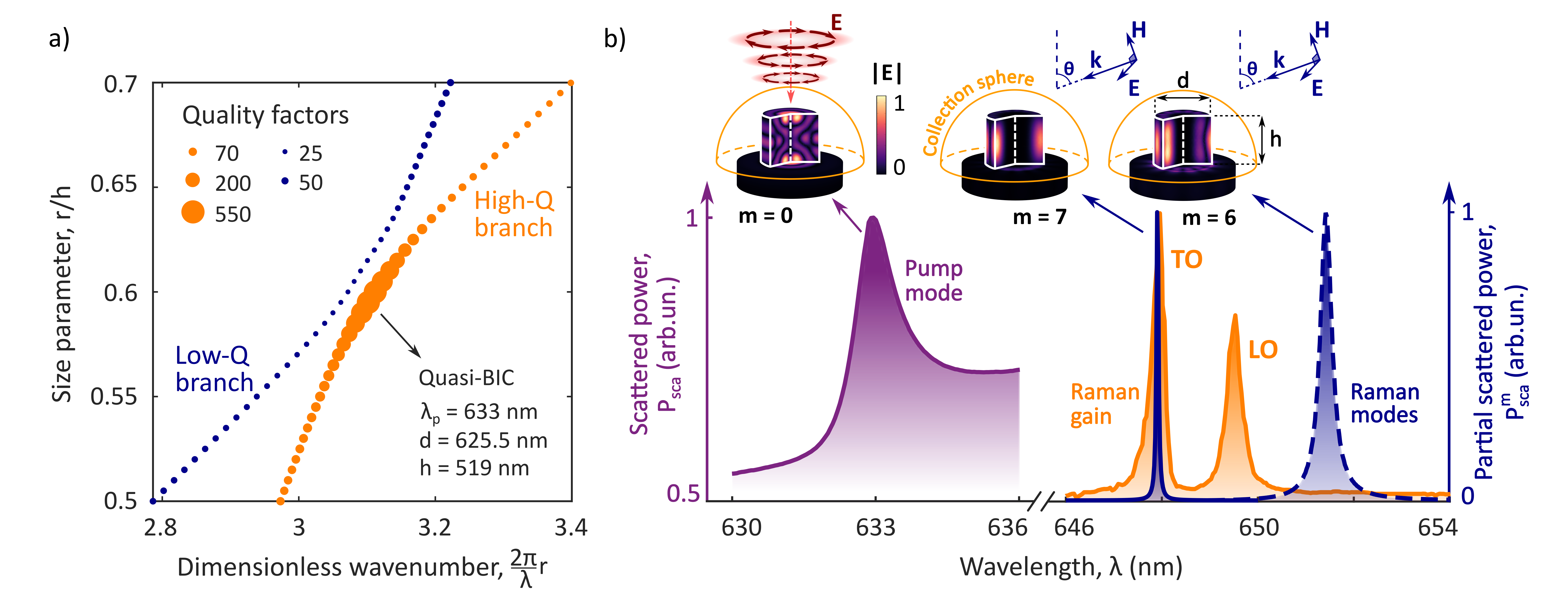}
\caption{\textbf{Excitation and emission modes optimization.} (a) The map of eigenmodes in dimensionless parameters of wavenumber $2\pi r/\lambda$ and size $r/h$, where $r = d/2$ is the cylindrical particle's radius, $h$ is its height and $\lambda$ is the free space eigenmode wavelength. The map is built for a gallium phosphide nanocylinder on a sapphire substrate and corresponds to the azimuthal number $m=0$. The size of the dots is proportional to the mode quality factor. The parameters of the quasi-BIC, manifesting itself in modes dispersions anticrossing behaviour, are shown specifically for pump wavelength $\lambda_{\mathrm{p}} = 633$~nm. (b) Dark violet curve in the shorter wavelength range depicts the scattered by quasi-BIC power $P_{\mathrm{sca}}$ spectrum for the case of azimuthally polarized vector beam excitation. In the longer wavelength range dark blue curves show the partial scattered power $P_{\mathrm{sca}}^{m}$ spectra corresponding to azimuthal numbers $m = 7$ and $m=6$ in case of TE-polarized plane wave incidence at an angle $\theta = 67^{\mathrm{o}}$ relative to the surface normal. The orange curve in the background shows spectral overlap of the Raman gain profile~\cite{Xiong:2003:1533-4880:335} with the found Raman modes. The insets show normalized electric field distributions and corresponding type of excitation. In all of the cases scattered radiation is collected into upper hemisphere.   \label{fig: quasi-BIC}}

\end{figure*}

The strong interaction of the modes inside the resonator leads to the appearance of high-$Q$ and low-$Q$ modes, whose dispersions in configuration space form an avoid crossing manifesting the Fridrich-Winteg BIC formation~\cite{bogdanov2019bound}. So, we search for the eigenmodes in the single cylindrical nanoparticle made of GaP on the sapphire ($\mathrm{Al_2O_3}$) substrate by using COMSOL Multiphysics~[Fig.~\ref{fig: quasi-BIC}(a)]. We specifically choose only the modes having rotational symmetry of electromagnetic field distribution with respect to the cylinder axis (azimuthal number $m = 0$) because this type of quasi-BICs usually demonstrates higher $Q$ factors and could be efficiently excited by azimuthal vector beam~\cite{koshelev2020subwavelength}. Gallium phosphide possesses negligible optical losses in the range of interest ($\lambda \sim 633-654$~nm covering both pump and Raman bands), so we build the map of eigenmodes in dimensionless coordinates which allows one to easily scale geometrical parameters of the resonator for desired wavelength. In particular, we choose the mode with highest quality factor on a high-Q dispersion branch and extract resonator's geometrical quantities, corresponding to the pump wavelength of interest $\lambda_{\mathrm{p}} = 633$~nm, from dimensionless wavenumber and size parameter. The resulting dimensions of the nanocylinder supporting quasi-BIC are diameter $d = 625.5$~nm, height $h = 519$~nm and related mode quality factor is $Q = 514$. The selected design is relatively sensitive to geometrical parameters deviations, e.g. 5~nm increase of the nanoparticle's diameter results in a 3~nm mode spectral shift, though still providing a high quality factor of $Q\sim420$. In order to prove that the found mode could be excited from the far-field we also plot scattered power spectrum for azimuthal vector beam excitation. It exhibits a clearly distinguishable peak with the background coming from a lower quality factor and longer wavelength mode [Fig.~\ref{fig: quasi-BIC}(b)].

One should expect that the efficiency of the nonlinear process in a single particle could be significantly increased by choosing a double-resonant structure with optical resonances spectrally splitted by the semiconductor phonon frequency similar to the principle of optical harmonics generation enhancement in nanostructures~\cite{thyagarajan2012enhanced, pashina2022thermo}. That is why, for the chosen configuration we also check the presence of the high-$Q$ Raman modes at the Stokes wavelength $\lambda_{\mathrm{S}}$. Though high-$Q$ modes usually have small scattering cross-section, we can still detect them in a lossless system by illuminating the nanoparticle with a planar wave at an angle relative to the surface normal, the same way as in dark-field microscopy. Figure~\ref{fig: quasi-BIC}(b) shows partially scattered power spectra $P_{\mathrm{sca}}^m$ for TE-polarized incident wave. They correspond to the WGMs whose fields possess specific cylindrical symmetry, meanwhile, the electric field has spatial dependence $\mathbf{E} \sim e^{im\varphi}$, where $m$ is the mode azimuthal number and $\varphi$ is the azimuthal angle with respect to the cylinder axis. Mode with $m = 7$ has perfect spectral overlap with transverse optical (TO) phonon peak providing narrow resonance line with FWHM $\approx0.06$~nm, which is less than that of a TO phonon with FWHM = 0.28~nm for specified pump wavelength~\cite{Xiong:2003:1533-4880:335}. Another high-$Q$ WGM with $m = 6$ is slightly detuned from the longitudinal optical (LO) peak and this spectral mismatch could be covered with a minor decrease of resonator diameter by $\sim2$~nm in a way to switch between two modes with high $Q$-factor. Otherwise, it was shown that even for non-perfect overlap, when the mode is located at the shoulder of the gain profile, a small seed of spontaneous radiation is sufficient to achieve lasing regime~\cite{takahashi2013micrometre}.

Raman emission is a nonlinear process and its enhancement requires simultaneous optimization for several parameters of the resonator. Therefore, we apply a simplified analytical model to predict the best configuration of the nanostructure. Lasing should occur when the round-trip gain exceeds the round-trip loss. The threshold could be then estimated as following~\cite{spillane2002ultralow}:

\begin{equation}
 P_{\mathrm{th}} = \frac{n_{\mathrm{S}}n_{\mathrm{p}}}{\lambda_{\mathrm{S}}\lambda_{\mathrm{p}}}\frac{V_{\mathrm{ eff}}}{Q_\mathrm{S}Q_\mathrm{p}}\frac{\pi^2}{g}\frac{1}{B\Gamma},
\label{eq:threshold}
\end{equation}
where $P_{\rm th}$ is the threshold power that differs from the one coupled to the resonator, $n_\mathrm{p}$ (or $n_\mathrm{S}$) is the index of refraction at the pump (or Stokes) wavelength, $V_{\rm eff}$ is the effective mode volume at Stokes wavelength, $g$ is the nonlinear Raman gain coefficient, $Q_\mathrm{p}$ (or $Q_\mathrm{S}$) is the total quality factor for the resonant mode excited at the pump (or Stokes) wavelength, $\Gamma$ is the spatial overlap factor of the modes and $B$ is the coupling efficiency between the mode and the excitation field.

The analysis of the modes' properties allows us to estimate the threshold for Raman lasing in the case of azimuthal vector beam excitation for both of the Raman modes (TO and LO) we found. For the Raman gain coefficient of GaP $g = 54$~cm/GW~\cite{678943} and along with the other introduced parameters (for details see \textbf{Supplemental document}), we obtain the following estimations for the thresholds: ${P_{\mathrm{th}}^{\mathrm{TO}} \approx 5}$~{mW}; ${P_{\mathrm{th}}^{\mathrm{LO}} \approx 35}$~{mW} corresponding to the moderate input power operational regime of the continuous HeNe laser. Therefore, we choose the introduced design to feed it to the rigorous model for Raman lasing.

\subsection{Rigorous model for Raman lasing}

\begin{figure*}[ht!]
\centering
\includegraphics[width=0.9\textwidth]{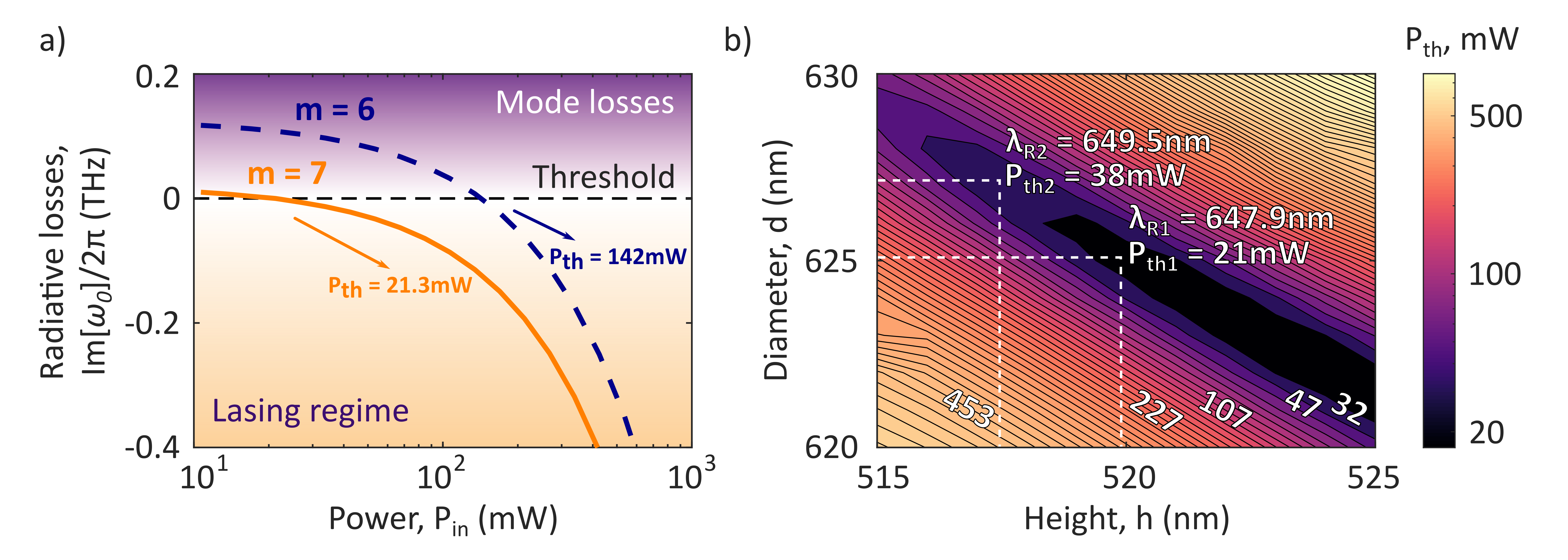}
\caption{\textbf{Raman lasing thresholds.} (a) Rigorous calculation of lasing thresholds for found design via full-wave simulations. The curves represent radiative losses $\operatorname{Im}[\omega_0]$ dependence from the pump power $P_{\mathrm{in}}$ for Raman modes with $m=7$ and $m=6$. The dashed horizontal line corresponds to the lasing threshold when the mode switches from losses to the lasing regime. (b) Isovalue map of the Raman thresholds for Raman mode with $m = 7$ obtained by a slight variation of size parameters. Perfect overlap cases of Raman mode with one of the gain peaks (at TO or LO phonon modes) are shown specifically with corresponding threshold values. \label{fig: threshold}}
\end{figure*}

In order to rigorously simulate Raman lasing threshold we utilize the wave equation for the Stokes component of the electric field $\mathbf{E}_{\rm S}$:
\begin{equation}
\Delta \mathbf{E}_{\rm S} + \mu_0\varepsilon_0 \varepsilon \omega_\mathrm{S}^2 \mathbf{E}_{\rm S} = -\mu_0 \omega_\mathrm{S}^2 \mathbf{P}_{\rm S},
\label{eq: wave_eq}
\end{equation} where $\mu_0$ and $\varepsilon_0$ are vacuum values of magnetic permeability and dielectric permittivity, $\omega_\mathrm{S} = 2\pi c/\lambda_{\mathrm{S}}$ is the frequency of Stokes photon and $\mathbf{P}_{\rm S}$ is the Raman polarization induced in the nanoparticle at the Stokes wavelength. For the low input intensities, Stokes polarization is defined mainly by spontaneous Raman scattering $\mathbf{P}_{\rm S} = \mathbf{P_{\rm sp}}(\omega_{\rm S}) = \hat{\alpha}(\omega_{\mathrm{S}}, \omega_{\mathrm{p}})\mathbf{E}_{\rm p}$, where $\hat{\alpha}$ is the Raman tensor and $\mathbf{E}_{\rm p}$ is the electric field excited at the pump frequency $\omega_{\mathrm{p}}=  2\pi c/\lambda_{\mathrm{p}}$. In nonlinear crystals spontaneous Raman seed emission $\mathbf{E}_{\rm S}$ could interact with the field at the pump frequency $\mathbf{E}_{\rm p}$ inducing stimulated contribution to Raman polarization:
\begin{equation}
    \mathbf{P}_{\rm st} = 6\varepsilon_0 \chi_{\mathrm{NL}}^{(3)}|\mathbf{E}_{\rm p}|^2 \mathbf{E}_{\rm S}.
\end{equation}
Here $\chi_{\mathrm{NL}}^{(3)}$ is the third order nonlinear susceptibility which is related to the Raman gain coefficient $g$~\cite{boyd2020nonlinear}:
\begin{equation} \label{eq: chi_nl}
    \begin{aligned}
        &\chi_{\mathrm{NL}}^{(3)} \sim \frac{1}{\omega_{\nu}^2 - \Omega^2 +2i\Omega\gamma }; \\
        &g=-\frac{3 \omega_{\mathrm{S}}}{n_{\mathrm{S}} n_{\mathrm{p}} c^2 \varepsilon_0} \operatorname{Im}\left[\chi_{\mathrm{NL}}^{(3)}\left(\omega_{\mathrm{S}}\right)\right],
    \end{aligned}
\end{equation}
where $\omega_{\nu}$ is the optical phonon frequency, $\Omega = \omega_{\mathrm{p}} - \omega_{\mathrm{S}}$ is a spectral detuning between pump and Stokes modes and $\gamma$ is a damping constant of the crystalline vibrational mode. Since we intentionally chose the modes frequencies in a way that $\omega_{\nu} = \Omega$, then nonlinear susceptibility is purely imaginary and negative and, by knowing Raman gain $g$, we can find susceptibility value to further implement it in our model. So, we substitute Stokes polarization in Eq.~(\ref{eq: wave_eq}) considering both spontaneous and stimulated contributions $\mathbf{P}_{\rm S} = \mathbf{P}_{\rm sp} + \mathbf{P}_{\rm st}$. The wave equation, therefore, takes the following form:
\begin{equation} \label{eq: modulation}
    \Delta \mathbf{E}_{\rm S} + \mu_0\varepsilon_0 \omega_\mathrm{s}^2 (\varepsilon + 6\chi_{NL}^{(3)} |\mathbf{E_{\mathrm{p}}(\mathbf{r})}|^2)  \mathbf{E}_{\rm S} = -\mu_0 \omega_\mathrm{s}^2 \hat{\alpha} \mathbf{E}_{\rm p}, 
\end{equation}
where dielectric permittivity is spatially modulated depending on the incident power and field distribution $|\mathbf{E_{\mathrm{p}}(\mathbf{r})}|^2$ excited at pump frequency $\omega_{\mathrm{p}}$. Since nonlinear susceptibility is purely imaginary and negative, this modulation reduces the overall losses of the system and gradual increase of the pump power should lead to generation rather than damping of Raman emission. This switching of the regimes indicates reaching the threshold and transition to the Raman lasing.  

By eliminating the source part in Eq.~(\ref{eq: modulation}), we can search for the complex eigenfrequencies $\omega_0$ of previously found TO and LO Raman modes as a function of incident power. To serve this purpose we use COMSOL Multiphysics and solve the problem utilizing two frequency domains. They correspond to the pump $\omega_{\mathrm{p}}$ and the Stokes $\omega_{\mathrm{S}}$ frequencies but operate within the same geometrical entities, which are GaP cylinders with previously defined diameter and height on a sapphire substrate surrounded by air. The quasi-infinite surroundings are mimicked by adding a perfectly matched layer at the outer boundaries of the simulation area which absorbs the incoming scattered radiation. The optical properties for gallium phosphide and sapphire are taken from Refs.~\cite{aspnes1983dielectric, dodge1986refractive}, correspondingly. In the first frequency domain, we calculate electromagnetic response for the found design in the case of azimuthally polarized vector beam excitation. We employ azimuthal symmetry of the introduced resonator and incident field to replace a fully three-dimensional problem with the axisymmetric two-dimensional one.  Furthermore, the amplitude of the induced electric field $\mathbf{E_{\mathrm{p}}(\mathbf{r})}$ is transferred to the second frequency domain to introduce spatial modulation for losses rate of Raman active medium: $\Delta\operatorname{Im}(\varepsilon) = 6\operatorname{Im}(\chi_{NL}^{(3)})|\mathbf{E_{\mathrm{p}}(\mathbf{r})}|^2$. Finally, we vary pump power of the incident beam  and solve the eigenproblem for such a perturbed resonator by choosing a proper field azimuthal number $m$. As a result, we plot the imaginary part of obtained eigenfrequencies, which consists of only the radiative part for the lossless resonator (zero extinction coefficient in the range of interest), and detect a switch to the lasing regime when this curve crosses zero [Fig.~\ref{fig: threshold}(a)]. The rigorous model provides us with higher threshold powers but is still achievable with a continuous wave laser for the TO mode with $m = 7$ which exhibits a threshold of $P_\mathrm{th} = 21.3$~mW. 

Finally, our full-wave approach allows us to estimate the thresholds for out-of-resonance conditions which is crucial for defining the required fabrication precision. By slightly varying the geometrical parameters of the cylindrical nanoparticle, we calculate corresponding thresholds for the Raman mode with $m=7$. The lasing threshold drastically increases when we retreat from quasi-BIC formation conditions [Fig.~\ref{fig: threshold}(b)], but could be still kept low if diameter and height are simultaneously adjusted in a way to stay on the dispersion branch of a high-$Q$ mode~[Fig.~\ref{fig: quasi-BIC}(a)]. For the latter case one needs to decrease the diameter of the resonator for increasing height which is shown with the low threshold area on the map [Fig.~\ref{fig: threshold}(b)].

Since the change in the resonator diameter causes also a slight shift in the resonant wavelength of Raman WGM (for details see \textbf{Supplemental document}), we specifically emphasize the parameters corresponding to the perfect overlap with one of the gain peaks (TO or LO). However, as was discussed earlier, even a small amount of spontaneous radiation would push the system out of unstable equilibrium near the threshold and, hence, the spectral shift of the Raman mode has less impact than non-efficient excitation at the pump frequency.

\section{Discussion}
The main idea of this work was to present the design of Raman nanolaser ready for experimental implementation which combines the advantages of low-threshold operation and subwavelength size. The proposed configuration is set up in a way to function most efficiently for the HeNe pump wavelength $\lambda_p = 633$~nm. Nevertheless, the concept of combining quasi-BIC excitation along with the emission from high-$Q$ mode could be transferred to a different wavelength or other Raman-active media. The major limitation lies in the trade-off between optical losses minimization for enhancement of modes quality factors and ensuring sufficient Raman gain values. The decrease in losses requires operation at lower frequencies, whereas Raman gain depends linearly on Stokes frequency [Eq.~(\ref{eq: chi_nl})] and, thus, favors higher input frequencies. Gallium phosphide appears to be the optimal material from this perspective since it possesses negligible optical losses in the visible range while still providing a high value of Raman gain.
\begin{figure}[t]
\includegraphics[width=\linewidth]{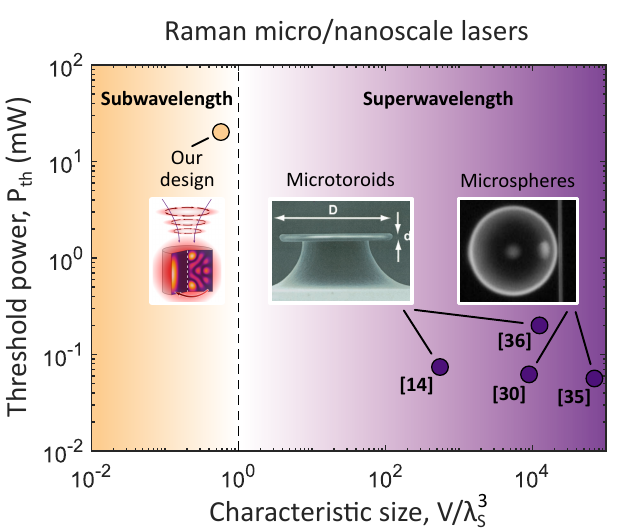}
\caption{\textbf{Comparison of micro/nanoscale Raman lasers.} The graph compares Raman lasers with dimensions not exceeding $50\;\mu$m in terms of their threshold power $P_\mathrm{th}$ and characteristic size $V/\lambda_{\mathrm{S}}^3$. Dark violet points correspond to previously reported superwavelength Raman lasers, which are toroidal and spherical cavities. Our design is shown with yellowish point and represents a separate category of subwavelength Raman lasers. 
\label{fig: comparison}}
\end{figure} 

Figure~\ref{fig: comparison} shows a brief comparison of the proposed design with existing analogues of Raman lasers, which dimensions do not exceed $50\;\mu$m, in terms of threshold power and characteristic size $V/\lambda_{\mathrm{S}}^3$. Such a comparison reveals a device scale as related to emission wavelength, which could be either subwavelength or superwavelength, and allows one to estimate power consumption. In Figure~\ref{fig: comparison}, we distinguish two categories of reported superwavelength Raman lasers, which are microspheres and microtoroids. Both of these types exploit WGMs as the main operating principle, where the relatively large radius of the devices ($\sim\!25\;\mu$m) enables excitation of the modes possessing high azimuthal orders and, consequently, huge quality factors ($Q \sim 10^8$). In accordance with the analytical model [Eq.~(\ref{eq:threshold})], it results in very low lasing thresholds $P_{\mathrm{th}}\approx60~\mu\mathrm{W}$ for both microspheres~\cite{spillane2002ultralow, min2003compact} and microtoroids~\cite{kippenberg2004ultralow}. The latter Raman microlaser utilizing microtoroids~\cite{shen2020raman} stands out of this row since even though it incorporates WGM excitation, light is inelastically scattered by the molecular vibrations of surface monolayer. So, the lasing is based on the surface stimulated Raman scattering and performs slightly higher threshold value $P_{\mathrm{th}} = 200~\mu\mathrm{W}$. Mentioned above configurations have extremely low lasing thresholds but this is achieved through the use of resonators that are significantly larger than the emission wavelength. Besides, both microspheres and microtoroids modes presented above are excited via tapered fiber, the gap between which and microresonator should be precisely controlled. This complicates the operation procedure as compared to the external excitation approach introduced in our work. Finally, we intentionally did not include in this graph Raman lasers based on photonic crystal cavities~\cite{takahashi2013micrometre, pradhan2019integrable} because, despite they have even lower threshold powers $P_{\mathrm{th}} \approx1~ \mu\mathrm{W}$ and small mode volumes, these cavities need a whole surrounding photonic crystal for operation and have larger footprint then.

\section{Conclusion}

We have proposed a design for a fully subwavelength Raman laser operating at a quasi-BIC possessing high $Q$-factor. Implementation of the quasi-BIC state in a high refractive index nanoparticle excited with the azimuthally polarized beam along with the utilization of high-$Q$ WGM for the Stokes line results in low-threshold Raman nanolaser. For design optimization, we introduce a double-resonant structure with the eigenmodes spectrally detuned from each other exactly by the TO phonon frequency. We apply a simple analytical model to optimize the geometrical parameters of the optical resonator and verify it with rigorous full-wave numerical analysis. In order to make our design feasible, we consider a gallium phosphide cylindrical nanoparticle with a diameter of about 0.95$\lambda$, where the lasing threshold has been found around 21~mW. Such conditions correspond to a continuous wave laser excitation with moderate intensity, which is promising for various practical applications related to optical sensing and frequency conversion. Our comparison with other designs has shown that the proposed ultrasmall Raman laser can be several orders of magnitude more compact.

\section{Acknowledgements}

This work was supported by the Ministry of Science and Higher Education of the Russian Federation (Project 075-15-2021-589) and Federal Academic Leadership Program Priority 2030.

\newpage

\bibliography{bibliography.bib}

\end{document}


\title{Supplementary Information for the Manuscript\\Subwavelength Raman Laser Driven by Quasi Bound State in the Continuum}

\author{Daniil Riabov}
\affiliation{ITMO University, Department of Physics and Engineering, Saint-Petersburg, Russia}

\author{Ruslan Gladkov}
\affiliation{ITMO University, Department of Physics and Engineering, Saint-Petersburg, Russia}

\author{Olesia Pashina}
\affiliation{ITMO University, Department of Physics and Engineering, Saint-Petersburg, Russia}

\author{Andrey Bogdanov}
\affiliation{ITMO University, Department of Physics and Engineering, Saint-Petersburg, Russia}
\affiliation{Qingdao Innovation and Development Center, Harbin Engineering University, Qingdao 266000, Shandong, China}

\author{Sergey Makarov*}
\affiliation{ITMO University, Department of Physics and Engineering, Saint-Petersburg, Russia}
\affiliation{Qingdao Innovation and Development Center, Harbin Engineering University, Qingdao 266000, Shandong, China}

\maketitle

\section{Analytical model for Raman lasing threshold}
To predict the best configuration of the nanostructure we firstly applied a simplified analytical model. The Raman lasing threshold could be then estimated as following~\cite{spillane2002ultralow}:

\begin{equation}
 P_{\mathrm{th}} = \frac{n_{\mathrm{s}}n_{\mathrm{p}}}{\lambda_{\mathrm{s}}\lambda_{\mathrm{p}}}\frac{V_{\mathrm{ eff}}}{Q_\mathrm{s}Q_\mathrm{p}}\frac{\pi^2}{g}\frac{1}{B\Gamma},
\label{eq:threshold_suppl}
\end{equation}
where $P_{\rm th}$ is the threshold power that differs from the one coupled to the resonator, $n_\mathrm{p}$ (or $n_\mathrm{S}$) is the index of refraction at the pump (or Stokes) wavelength~\cite{aspnes1983dielectric}, $V_{\rm eff}$ is the effective mode volume at Stokes wavelength, $g$ is the nonlinear Raman gain coefficient, $Q_\mathrm{p}$ (or $Q_\mathrm{S}$) is the total quality factor for the resonant mode excited at the pump (or Stokes) wavelength, $\Gamma$ is the spatial overlap factor of the modes and $B$ is the coupling efficiency between the mode and the excitation field.

Simplified analytical model for Raman lasing threshold calculation in Eq.~(\ref{eq:threshold_suppl}) requires calculation of pump and Stokes modes' parameters. We start from effective mode volume which has a meaning of substitute for a nontrivial spatial field distribution with homogeneous pattern but of a smaller volume. The exact expression for effective mode volume is~\cite{srinivasan2006cavity}:

\begin{equation}
\label{eq:veff}
V_{\mathrm{eff}}=\frac{\int_{V} \varepsilon(\mathbf{r})|\mathbf{E}(\mathbf{r})|^{2} d^{3} \mathbf{r}}{\max \left[\varepsilon(\mathbf{r})|\mathbf{E}(\mathbf{r})|^{2}\right]}
\end{equation}
where $\mathbf{E}(\mathbf{r})$ is the electric field of the mode, $V$ is the volume of the particle and $\varepsilon(\mathbf{r})$ is the relative permittivity of the material. Since the dielectric permittivity is considered to be homogeneous over the particle volume it could be excluded from the calculations. For introduced high-$Q$ WGMs at the Stokes frequency we find effective mode volumes to be $V_{\mathrm{eff}}^{m = 7} = 0.105V$ and $V_{\mathrm{eff}}^{m = 6} = 0.2V$.  

For estimation of the threshold power $P_{\mathrm{th}}$ one needs also to calculate the overlap factor~\cite{zograf2020stimulated}:
\begin{equation}
\label{eq:overlap}
\Gamma = \frac{1}{V} \int_V \frac{\left|{\mathrm{\mathbf{E}}_{\mathrm{p}}}(\mathbf{r})\cdot {\mathrm{\mathbf{E}}_{\mathrm{S}}}(\mathbf{r}) \right|}{\left|{\mathrm{\mathbf{E}}_{\mathrm{p}}}(\mathbf{r})\right|\cdot \left|{\mathrm{\mathbf{E}}_{\mathrm{S}}}(\mathbf{r})\right|} d^{3} \mathbf{r}
\end{equation}
where ${\mathrm{\mathbf{E}}_{\mathrm{p}}}(\mathbf{r})$ and ${\mathrm{\mathbf{E}}_{\mathrm{S}}}(\mathbf{r})$ are the electric fields of the eigenmodes corresponding to pump and Stokes photon wavelengths, respectively. This factor could take the values from 0 to 1 depending on how well mode spatial distributions intersect with each other. The closer this value to 1 the lower resulting lasing threshold is, since the modes should overlap well in order to transfer energy via electron-phonon interaction. For the modes in the main text the values of overlap factors are $\Gamma^{m = 7} = 0.308$ and $\Gamma^{m = 6} = 0.477$.

In analogy to the previous formula, we estimate coupling efficiency of incident field with the optical mode at the pump frequency:

\begin{equation}
    B = \frac{1}{V} \int_V \frac{\left|{\mathrm{\mathbf{E}}_P}(\mathbf{r})\cdot {\mathrm{\mathbf{E}}_{\mathrm{inc}}}(\mathbf{r}) \right|}{\left|{\mathrm{\mathbf{E}}_P}(\mathbf{r})\right|\cdot \left|{\mathrm{\mathbf{E}}_{\mathrm{inc}}  }(\mathbf{r})\right|} d^{3} \mathbf{r},
\end{equation}
where $\mathbf{E}_{\mathrm{inc}}(\mathbf{r})$ is the excitation field amplitude. Since the found quasi-BIC possesses only components of the field along $\mathbf{e}_{\varphi}$ unit vector in cylindrical coordinate system, the mode field is always aligned with azimuthal beam electric vector which provides the highest value for coupling efficiency estimation $B = 1$.

For mode quality factors calculation we utilize COMSOL Multiphysics eigensolver. Since geometry of the resonator has rotational symmetry we replace fully three-dimensional problem with axisymmetric two-dimensional one. By stating specifically the azimuthal number for the mode field we find corresponding quality factors to be $Q_{S}^{m = 7} = 10977$ and $Q_{S}^{m = 6} = 1818$. 

\section{Raman WGM dependence on geometrical parameters variation}
\begin{figure}[htbp]
\centering
\fbox{\includegraphics[width=.6\linewidth]{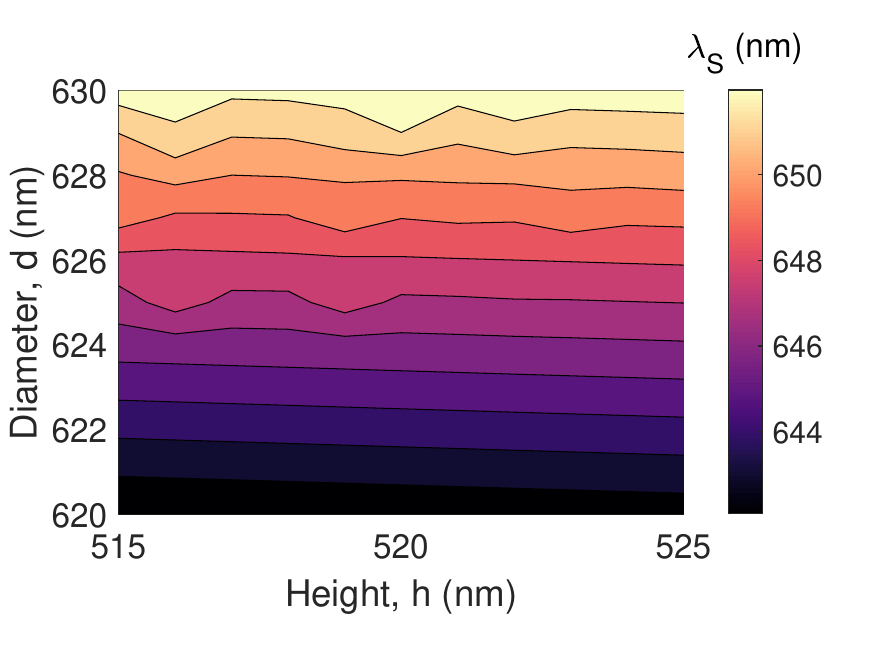}}
\caption{The isovalue map of the resonant wavelength $\lambda_{\mathrm{S}}$ for introduced mode with $m=7$ as a function of nanocylinder's diameter and height.}
\label{fig:Ram_wvln}
\end{figure}

We utilized our full-wave numerical approach to estimate the thresholds for out-of-resonance conditions which is crucial for defining the required fabrication precision. By slightly varying the geometrical parameters of the cylindrical nanoparticle, we calculated corresponding thresholds for the Raman WGM with $m=7$. However, this variation also causes slight shift of the resonant wavelength and, therefore we plot eigenmode wavelength dependence on the height and diameter of the nanocylinder [Fig.~\ref{fig:Ram_wvln}].

From the Figure~\ref{fig:Ram_wvln} it is clear that the introduced mode is indeed WGM since its resonant wavelength strongly depends on nanoparticle's diameter but is not sensitive to the height variations.

\bibliography{bibliography_suppl.bib}